\documentclass[aps,prl,twocolumn,superscriptaddress,showpacs]{revtex4}
\usepackage{graphicx}

\begin{document}
\title{Universal quantized spin-Hall conductance fluctuation in graphene}
\author{Zhenhua Qiao}
\affiliation{Department of
Physics and the center of theoretical and computational physics, The
University of Hong Kong, Hong Kong, China}
\author{Jian Wang$^*$}
\affiliation{Department of Physics and the center of theoretical and
computational physics, The University of Hong Kong, Hong Kong,
China}
\author{Yadong Wei}
\affiliation{Department of Physics, School of physics, Shenzhen
University, Shenzhen, China}
\author{Hong Guo}
\affiliation{Department of Physics, McGill University, Montreal,
PQ, Canada H3A 2T8}

\begin{abstract}
We report a theoretical investigation of quantized spin-Hall
conductance fluctuation of graphene devices in the diffusive regime.
Two graphene models that exhibit quantized spin-Hall effect (QSHE)
are analyzed. Model-I is with unitary symmetry under an external
magnetic field $B\ne 0$ but with zero spin-orbit interaction,
$t_{SO}=0$. Model-II is with symplectic symmetry where $B=0$ but
$t_{SO} \ne 0$. Extensive numerical calculations indicate that the
two models have exactly the same universal QSHE conductance
fluctuation value $0.285 e/4\pi$ regardless of the symmetry.
Qualitatively different from the conventional charge and spin
universal conductance distributions, in the presence of edge states
the spin-Hall conductance shows an one-sided log-normal distribution
rather than a Gaussian distribution. Our results strongly suggest
that the quantized spin-Hall conductance fluctuation belongs to a
new universality class.
\end{abstract}
\maketitle

One of the most important transport features of mesoscopic
conductors is the \emph{universal} conductance fluctuation (UCF) in
the diffusive regime caused by disorder scattering and quantum
coherence\cite{lee85}. The universality characterized by the value
of UCF only depends on the dimensionality and symmetry of the
system. According to random matrix theory (RMT)\cite{beenakker},
there are three ensembles or universalities due to symmetry: (1)
when time-reversal and spin-rotation symmetries are present,
\emph{i.e.} when magnetic field $B=0$ and spin-orbit interaction
(SOI) $t_{SO}=0$, the Hamiltonian $H$ of the system is an orthogonal
matrix and one has circular orthogonal ensemble (COE). COE is
characterized by a symmetry index $\beta=1$. (2) If time-reversal
symmetry is broken by $B\neq 0$, $H$ is unitary and one has the
circular unitary ensemble (CUE) characterized by $\beta=2$. (3)If
spin-rotation symmetry is broken by $t_{SO}\neq 0$ while
time-reversal symmetry is maintained, one has the circular
symplectic ensemble (CSE) for which $\beta=4$. While different
ensembles have different values of UCF, it is amazing that the
multitudes possibilities of electron dynamics in nature can be
classified by only a few ensembles\cite{foot1}. For instance, in one
dimension (1D) the UCF value is given by\cite{beenakker}
$[rms(G)]^2=2/(15\beta)$.

Recently, \emph{universal} fluctuation was also found to occur in 2D
mesoscopic spin-Hall effect (SHE)\cite{ren}. SHE can be induced by
spin-orbit interaction, for instance Rashba SOI in 2D, such that
chemical potentials of the spin-up or -down channels become
different at the two boundaries of a mesoscopic
sample\cite{murakami,sinova}. With disorder, numerical calculations
showed\cite{ren} that the spin-Hall conductance $G_{sH}$ of a 2D
mesoscopic system fluctuates from sample to sample with a value
$rms(G_{sH})\approx 0.18 e/4\pi$: this is independent of system
details thus universal, and the phenomenon is termed universal
spin-Hall conductance fluctuation (USCF). The numerical value of
USCF has been quantitatively confirmed by RMT\cite{confirm}. For
most situations, $G_{sH}$ itself may have any value in units of
$e/4\pi$ depending on system details. On the other hand, several
authors have advanced the notion of \emph{quantized} SHE (QSHE) for
situations where electronic edge states exist: in QSHE $G_{sH}$
takes integer multiples of $e/4\pi$. In particular, QSHE is shown to
occur in 2D graphene due to SOI plus the peculiarity of graphene
electronic structure\cite{kane}. QSHE is also predicted to occur in
graphene without SOI but with an external magnetic
field\cite{palee}. Therefore, using the language of
RMT\cite{beenakker}, QSHE occurs in graphene with CUE where $B\neq
0$ but $t_{SO}=0$; and with CSE where $B=0$ but $t_{SO}\neq 0$.

Several important and interesting questions therefore arise
concerning the universality of QSHE: is it still classifiable by the
RMT ensembles? As the disorder is increased, is there a USCF for
QSHE and if there is, is the value different from the USCF for SHE
that is $0.18e/4\pi$? What is the distribution of $G_{sH}$ in QSHE?
Indeed, all these questions are related to the curiosity,
\emph{i.e.} whether or not the Dirac dispersion relation of graphene
brings new physics to the spin-Hall conductance fluctuation in the
quantized SHE. It is the purpose of this work to investigate these
issues.

To be more specific, we investigate the two graphene models that
exhibit QSHE\cite{palee,kane} as mentioned above. In the first
model, model-I\cite{palee}, SOI is neglected in the graphene but a
magnetic field is applied causing a Zeeman splitting. Model-I has
unitary symmetry and importantly is in the quantum Hall regime where
edge states are present.
Due to the Zeeman splitting and graphene energy spectrum both
electron-like and hole-like edge states exist near the Fermi level
forming counter-circulating edge states in graphene that has been
confirmed experimentally\cite{palee1}. It is these
counter-circulating edge states that lead to QSHE\cite{palee}. The
second model, model-II, is the one proposed by Kane and
Mele\cite{kane} where intrinsic SOI gives rise to "spin filtered"
edge states that cause QSHE based on an idea discussed by
Haldane\cite{haldane}. Clearly, model-II has symplectic symmetry.
Although the value of SOI parameter $t_{SO}$ for graphene is
small\cite{macdonald}, model-II is nevertheless very useful for our
purpose, namely to investigate universality class of QSHE. As we
show later, the value of $t_{SO}$--as long as it is nonzero, turns
out to be irrelevant as far as universality is concerned. From the
symmetry point of view, one would expect these two models to belong
to different universality classes. To our surprise, extensive
numerical results indicate that in the presence of edge states, the
QSHE dominates the physics and these two models give exactly the
same universal value for ${\rm USCF} = 0.285 e/4\pi$ regardless of
symmetry. The distribution of $G_{sH}$ in the QSHE regime is found
to obey an one-sided log-normal distribution: this is qualitatively
different from the conventional UCF for charge and USCF for SHE
where it is a Gaussian distribution.

\begin{figure}
\includegraphics[width=8cm,totalheight=6.5cm,angle=0]{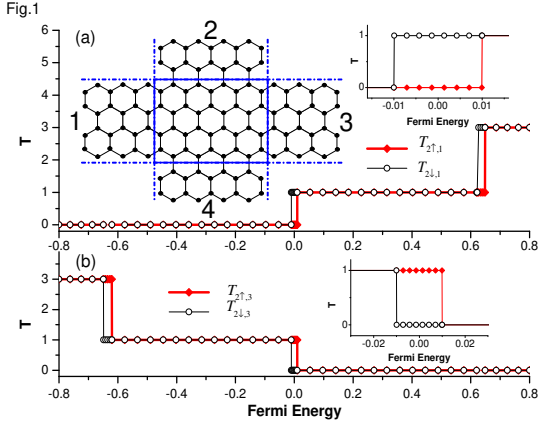}
\caption{(Color online) The transmission coefficient $T_{21}$ and
$T_{23}$ versus energy at a fixed magnetic flux. Inset: schematic
plot of the four terminal mesoscopic sample where the intrinsic SO
interaction exists in the center scattering region and the leads
$1$, $3$. And the Rashba SO only exists in the center part and the
leads $1$, $3$, when the spin-Hall conductance is measured through
leads $2$, $4$.} \label{fig1}
\end{figure}

In a tight-binding representation, the Hamiltonian for 2D honeycomb
lattice of graphene can be written as:
\begin{eqnarray}
H_1&=&\sum_{i \sigma} {\epsilon_{i} c^\dagger_{i \sigma} c_{i
\sigma}} -t \sum_{<ij>\sigma}{e^{i{2\pi}
{\phi}_{ij}}}c^\dagger_{i\sigma}c_{j\sigma}
\nonumber \\
&&+g_s \sum_{i \sigma } c^\dagger_{i\sigma} ({\bf \sigma} \cdot {\bf
B}) c_{i\sigma}
\end{eqnarray}
for model-I, and
\begin{eqnarray}
H_2&=&\sum_{i \sigma} {\epsilon_{i} c^\dagger_{i \sigma} c_{i
\sigma}}-t \sum_{<ij>\sigma}c^\dagger_{i\sigma}c_{j\sigma}
\nonumber \\
&+&\frac{2i}{\sqrt{3}}t_{SO}\sum_{\ll{ij}\gg}
{c^\dagger_{i}{\sigma}{\cdot}(\mathbf{d}_{kj}{\times}\mathbf{d}_{ik})c_{j}}
\end{eqnarray}
for model-II, where $c^\dag_{i\sigma}$ ($c_{i}$) is the creation
(annihilation) operator for an electron with spin $\sigma$ on site
${i}$. The first term in $H_1$ and $H_2$ is the on-site single
particle energy where diagonal disorder is introduced by drawing
$\epsilon_{i}$ randomly from a uniformly distribution in the
interval $[-W/2,W/2]$. Here $W$ measures strength of disorder. The
second term in $H_1$ is due to nearest neighbor hopping and the
presence of a magnetic field, the last term in $H_1$ is due to
Zeeman energy. Here $g_s=(1/2)g\mu_B$ (with $g=4$) is the Lande g
factor, phase ${\phi_{ij}=\int{\bf A}{\cdot}d{l}/\phi_{0}}$,
$\phi_{0}=h/e$ is the quantum of flux, and the spin-Hall conductance
and its fluctuation are in unit of $e/4\pi$.
In $H_2$ the last term is the SOI that involves next nearest sites
of indices $i,j$ with $k$ the common nearest neighbor of $i$ and
$j$, and ${\mathbf{d}}_{ik}$ describes a vector pointing from $k$ to
$i$.

We use the four-probe device schematically shown in the inset of
Fig.1 to investigate USCF in QSHE. The four probes are exact
extensions from the central scattering region, \emph{i.e}. the
probes are graphene nano-ribbons. The number of sites in the
scattering region is denoted as $N=n_x{\times}n_y$, where there are
$n_x=8{\times}n+1$ sites on $n_y=4{\times}n$ chains. We apply
external bias voltages $V_i$ with $i=1,2,3,4$ at the four different
probes as $V_{i}=(v/2,0,-v/2,0)$. The spin-Hall and charge Hall
conductance $G_{sH}$ and $G_H$ can be calculated from the
multi-probe Landauer-Buttiker formula\cite{ren}:
\begin{eqnarray}
G_{sH}&=&(e/8{\pi})[(T_{2{\uparrow},1}-T_{2{\downarrow},1})-(T_{2{\uparrow},3}-T_{2{\downarrow},3})]
\nonumber \\
G_{H}&=&(e^2/h)[(T_{2{\uparrow},1}+T_{2{\downarrow},1})-(T_{2{\uparrow},3}+T_{2{\downarrow},3})]
\label{eq2}
\end{eqnarray}
where the transmission coefficient is given by
$T_{2{\sigma,1}}=Tr(\Gamma_{2{\sigma}}G^{r}\Gamma_{1}G^{a})$ with
$G^{r,a}$ being the retarded and advanced Green functions of the
central disordered region which can be evaluated numerically. The
quantities $\Gamma_{i{\sigma}}$ are the linewidth functions
describing coupling of the probes and the scattering region and are
obtained by calculating self-energies $\Sigma^r$ of the
semi-infinite leads using a transfer matrices method\cite{lopez84}.
The spin-Hall conductance fluctuation is defined as
$\text{rms}(G_{sH})\equiv \sqrt{\left\langle G_{sH}^{2}\right\rangle
-\left\langle G_{sH}\right\rangle ^{2}}$, where $\left\langle
{\cdots}\right\rangle $ denotes averaging over an ensemble of
samples with different configurations of the same disorder strength
$W$. In the following, our numerical data are mainly collected on a
system with $n=8$, \emph{i.e.} with ${32\times}65$ sites in the
graphene. In the rest of the paper, we fix units by setting energy
$E$, disorder strength $W$, SOI coupling $t_{SO}$ in terms of the
hopping parameter $t$, and the magnetic field in terms of magnetic
flux $\phi$.

\begin{figure}
\includegraphics[width=8cm,totalheight=6.5cm,angle=0]{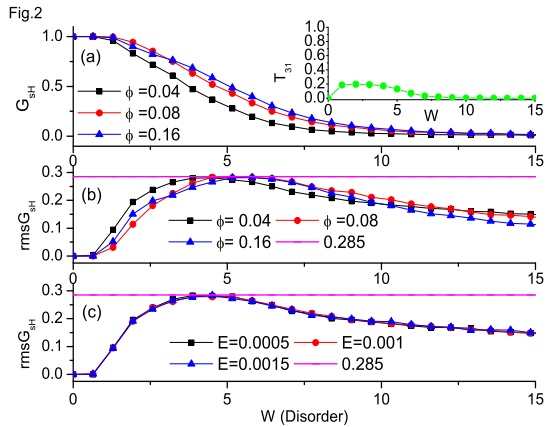}
\caption{(Color online) Spin-Hall conductance and its fluctuation
versus disorder strength at different energies and magnetic fluxes
for the first model. Inset: the transmission coefficient $T_{31}$
versus disorders. } \label{fig2}
\end{figure}

We first examine model-I which has unitary symmetry. Fig.\ref{fig1}
shows the transmission coefficient $T_{2\sigma,1}$ and
$T_{2\sigma,3}$ as a function of energy with $\phi=3\sqrt{3}/128$
and without disorder, the spin index $\sigma=\uparrow,\downarrow$.
We observe that if we neglect the Zeeman energy the quantum charge
Hall conductance takes the well known result\cite{zhang} $G_H=\pm
4(|n|+1/2)(e^2/h)$. In addition, because of edge states we see that
$T_{21}$ is nonzero and $T_{23}$ is zero above the Fermi level
$E=0$, while $T_{21}$ is zero and $T_{23}$ is nonzero below Fermi
level exhibiting hole-like behavior. Due to the Zeeman shift we have
$T_{2{\uparrow},1}=T_{2{\downarrow},3}=0$ and
$T_{2{\downarrow},3}=T_{2{\uparrow},1} \ne 0$ near Fermi level. From
Eq.(\ref{eq2}) we obtain QSHE: $G_{sH}=1$ in unit of $e/4\pi$ and
$G_H=0$. Here we emphasize that the bias voltage is applied across
probes 1 to 3 (see Fig\ref{fig1}) and it causes a transverse flow of
spin-current between probes 2 and 4 that leads to the QSHE.

In the regime of QSHE, we now increase disorder strength $W$. This
causes a break down of the integer value of $G_{sH}$ and induces
sample to sample fluctuations of $G_{sH}$. Fig.\ref{fig2}a plots the
average $G_{sH}$ by calculating 5000 samples for each point on the
figure, Fig.\ref{fig2}b plots the corresponding fluctuation
$rms(G_{sH})$, as a function of $W$. When $W$ is increased, $G_{sH}$
decreases from its quantized value $G_{sH}=1$ and $rms(G_{sH})$
increases. The break down of quantized $G_{sH}$ is due to $W$ that
causes a direct transmission from probe 1 to 3 (see Fig.\ref{fig1}),
this is shown in the inset of Fig.\ref{fig2}a where the direct
transmission $T_{31}$ is plotted against $W$. From $T_{31}$ we
conclude that the graphene device is in an insulating regime at
small $W$, \emph{i.e.} zero or very small $T_{31}$; it is in a
diffusive regime for intermediate $W$ and finally reentrant to the
insulating regime for large $W$. For a given $E$ or $\phi$,
$rms(G_{sH})$ develops a ``plateau" region, \emph{e.g.} in the range
$W=[3,7]$ in Fig.\ref{fig2}b. This plateau is at $rms(G_{sH})=0.285$
in unit of $e/4\pi$. The plateau range of $W$ depends on specific
values of $E$ or $\phi$, but we found $rms(G_{sH})=0.285$ is always
true if there is a plateau, \emph{i.e.}, if the diffusive transport
regime is established. We therefore identify $rms(G_{sH})=0.285$ as
a ``universal" value. This USCF value is different from that of the
conventional SHE situation\cite{ren,confirm} where the universal
value is $0.18$. Therefore QSHE and SHE belong to different
universality classes due to this different statistical property.

\begin{figure}
\includegraphics[width=8cm,totalheight=6.5cm,angle=0]{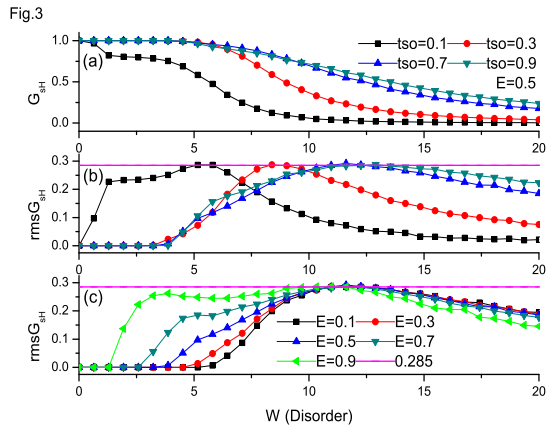}
\caption{(Color online) Spin-Hall conductance and its fluctuation
versus disorder strength at different energies and magnetic fluxes
for the second model. The parameters used in (a) and (b) are the
same.} \label{fig3}
\end{figure}

Next, we investigate Model-II that has a symplectic symmetry. For
such a graphene device there is an energy gap between $-1 < E <1$,
within which edge states exist\cite{kane}. Fig.3 plots averaged
$G_{sH}$ and $rms(G_{sH})$ versus $W$ for a given set of $E,\
t_{SO}$ parameter values. 5000 samples were calculated for the
disorder averaging. Similar behavior is found as that of Model-I.
For different values of $t_{SO}$, $rms(G_{sH})$ reaches a plateau at
different range of $W$ (see Fig.\ref{fig3}). Amazingly, all plateaus
have the same value and this value is precisely $rms(G_{sH})=0.285$!
To further confirm this finding, Fig.3c plots $rms(G_{sH})$ vs $W$
for a fixed $t_{SO}$ but several different values of energy $E$.
Again, same conclusion is obtained. This indicates that there exist
a transport regime where the QSHE conductance fluctuation has a
universal behavior independent of disorder (albeit a narrow region),
energy and SOI. Results of Fig.\ref{fig2} and Fig.\ref{fig3}
strongly suggest that there is a universal spin-Hall conductance
fluctuation in the quantized spin-Hall regime with USCF$=0.285$ in
unit of $e/4\pi$. This is different from the conventional SOI
induced SHE where USCF$=0.18$\cite{ren}.

Very importantly, it appears that symmetry does not play a role in
the QSHE regime at least for the CUE and CSE cases we have examined:
both give $rms(G_{sH})=0.285$. To further support this finding, we
calculated the distribution function of $G_{sH}$, $P(G_{sH})$, in
the QSHE regime. Such a distribution is a Gaussion for conventional
SHE in the diffusive regime\cite{ren}. For QSHE, Fig.4a-d plot
$P(G_{sH})$ for four different values of $W$ in the universal regime
for Model-II which has CSE symmetry. Data were collected by
calculating 84,000 samples for each $W$. The distributions are
completely different from a Gaussian! We found that by using
$ln(G_{sH})$ as a variable and plot $P(ln(G_{sH}))$, all the
distributions become one-sided log-normal (see Fig.4e-h). For
Model-I which has CUE symmetry, our results show the same
conclusion, \emph{i.e.} the distribution of quantum spin-Hall
conductance is an one-sided log-normal. Therefore, for the two
models we investigated, not only USCF $rms(G_{sH})=0.285$ is the
same, but also the distribution function is the same. This strongly
indicates that in the presence of edge states (\emph{i.e.} QSHE),
systems with unitary symmetry and symplectic symmetry belong to the
same universality class that is different from the conventional SHE.

\begin{figure}
\includegraphics[width=8cm,totalheight=6.5cm,angle=0]{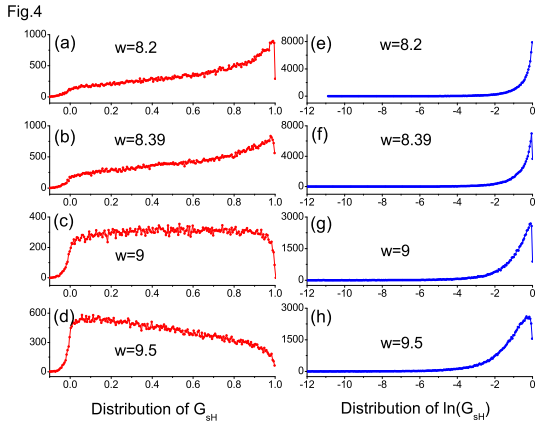}
\caption{(Color online)(a)-(d) The distribution of spin-Hall
conductance at different disorder strengths for the second model.
(e)-(f) The distribution of $ln(G_{sH})$. } \label{fig4}
\end{figure}

Finally, as a further confirmation of the QSHE universality class,
we have carried out extensive calculation on spin-Hall conductance
fluctuation for the same four probe graphene device with additional
Rashba SOI $t_R$\cite{sheng}. For {\it non-zero} $t_R$, three cases
are of interest. (1). $B=0$ and $t_{SO}=0$. For this situation it is
obvious that there is no edge state and therefore spin Hall effect
caused by $t_R$ is not quantized. Indeed, here we did not obtain the
USCF for QSHE but obtained a value of $0.18$ for all energies,
\emph{i.e.}, the same as the conventional USCF found
before\cite{ren,confirm}. As expected, for this case the
distribution of $G_{sH}$ was found to be a Gaussian. (2). When
$|E|<1$, for both model I and model II our numerical results show
that ${\rm USCF}=0.285$ remains the same as long as $t_R$ does not
destroy the edge states. (3). When $|E|>1$, there is no edge states
in model II\cite{kane}, our results show that ${\rm USCF} = 0.18$
for any $t_{SO}$. Therefore, edge states dominate the quantized
spin-Hall physics and $t_R$ is an irrelevant parameter (for both
model I and model II). On the other hand, if edge states are absent
$t_{SO}$ becomes an irrelevant parameter (for model II). This
clearly shows the landscape of universality class and it is the edge
state that drives the system from the universality of ${\rm
USCF}=0.18$ to the new universality we have discussed.

In summary, we have investigated quantized spin-Hall conductance
fluctuation for two models with unitary and symplectic symmetry,
respectively. Our numerical results show that both models exhibit
the same universal quantum spin-Hall conductance fluctuation with
the value $0.285 e/4\pi$. Due to the presence of edge states, the
distribution of quantum spin-Hall conductance obeys one-sided
log-normal distribution for both models. This strongly suggests that
the quantized spin-Hall conductance fluctuation for systems with
both unitary symmetry and symplectic symmetry belong to the same
universality class that is different from the usual spin-Hall
conductance fluctuation in the absence of edge states.

\bigskip

\section{acknowledgments}
This work was financially supported by a RGC grant (HKU 704607P)
from the government of HKSAR and LuXin Energy Group (J.W), NSFC
under grant No. 10574093 (Y.D.W),  and NSERC of Canada, FQRNT of
Qu\'{e}bec and Canadian Institute of Advanced Research (H.G). We
thank Dr. W. Ren for useful discussions on the analysis of the log
normal distributions.

$^*$ Electronic address: jianwang@hkusua.hku.hk


\begin{thebibliography}{99}
\bibitem{foot1}
An exceptional case to the COE, CUE and CSE universalities is in the
integer quantum Hall regime where it is the edge states that carry
charge-current. As disorder is increased and quantized Hall plateaus
destroyed, the Hall conductance fluctuates from sample to sample.
The universality class of Hall conductance fluctuation in the
quantum Hall regime has not been well understood.
\end{thebibliography}
\end{document}